\documentclass[conference]{IEEEtran}
\usepackage{cite}
\usepackage{amsmath,amssymb,amsfonts}
\usepackage{algorithmic}
\usepackage{url}
\usepackage{graphicx}
\usepackage{textcomp}
\usepackage{xcolor}
\def\BibTeX{{\rm B\kern-.05em{\sc i\kern-.025em b}\kern-.08em
    T\kern-.1667em\lower.7ex\hbox{E}\kern-.125emX}}
\begin{document}

\title{Emovectors: assessing emotional content in jazz improvisations for creativity evaluation
}

\author{\IEEEauthorblockN{Anna Jordanous}
\IEEEauthorblockA{\textit{School of Computing} \\
\textit{University of Kent}\\
Canterbury, UK\\
a.k.jordanous@kent.ac.uk}
}

\maketitle


\begin{abstract}
  Music improvisation is fascinating to study, being essentially a live demonstration of a creative process. In jazz, musicians often improvise across predefined chord progressions (leadsheets). How do we assess the creativity of jazz improvisations? And can we capture this in automated metrics for creativity for current LLM-based generative systems? Demonstration of emotional involvement is closely linked with creativity in improvisation. Analysing musical audio, can we detect emotional involvement? This study hypothesises that if an improvisation contains more evidence of emotion-laden content, it is more likely to be recognised as creative. An embeddings-based method is proposed for capturing the emotional content in musical improvisations, using a psychologically-grounded classification of musical characteristics associated with emotions. Resulting `emovectors' are analysed to test the above hypothesis, comparing across multiple improvisations. Capturing emotional content in this quantifiable way can contribute towards new metrics for creativity evaluation that can be applied at scale.
\end{abstract}


\section{Introduction}

Music improvisation is fascinating to study for musical creativity, as it is a domain which is inherently the live demonstration of a creative process as it happens. A common scenario in jazz improvisation is where a group of musicians play together through multiple repeats of a set of predefined chords (a leadsheet) for a song or melody, typically with one musician soloing (playing an improvisation) at a time, and the other musicians playing accompanying backings to that solo. 

How do we detect or evaluate more creative jazz improvisations, particularly through computational metrics? Automated metrics for creativity, that can scale up to analysis of Big Data, would be particularly valuable in assessing and guiding output of current LLM-based music AI systems, if they can be grounded in understanding of what it means to be creative as an improviser. One area with strong links towards enhanced perceived creativity is the demonstration of emotional involvement by the improvising musician \cite{jordanous12jims}. It is recognised that music content can express emotions \cite{sloboda2001emotions, juslin2013does}. In scenarios where the only data we have available for analysis is music audio data such as music recordings, we can detect indicators of emotional content within that audio data that can be treated as proxy evidence for emotional involvement.

This study investigates the research hypothesis that: if an improvisation contains more evidence of emotion-laden content, it is more likely to be recognised as creative. 

An embeddings-based method is proposed to capture the emotional content in music improvisations, using a psychologically-grounded classification of musical characteristics associated with emotions to generate emotion embeddings from audio features \cite{juslin03}. The resulting embeddings, or ‘emovectors’, are analysed to test the above hypothesis across multiple improvisations. Results so far suggest this approach could provide a method to capture and measure emotional content, towards metrics for creativity evaluation at scale.


\section{Background}

This work is motivated by the need for automated tests for music improvisation creativity, which can be applicable for large data. Tools for generating music and audio have been an area of research for decades e.g. \cite{Allan2004, keller2010}. Older tools typically could generate output at a reasonable scale; however the advent of recent LLM-based music generative AI tools such as MusicGen \cite{copet2024simplecontrollablemusicgeneration} has transformed the scale at which credible outputs can be generated. With greater scale comes greater difficulties for any evaluative processes which require human input or more lengthy, involved tests which cannot be applied practically at scale. For example, evaluating the creativity of the output of such music generators is a complex and unsolved problem \cite{10.1145/2967506}, typically relying on tests which require manual intervention, interactivity (causing a fitness bottleneck when applied at scale), and/or application of automated tests that do not necessarily correlate with what is important for musical creativity itself. This is driven by pragmatism; creativity is challenging to define and test for. 

A `standard definition of creativity', the combination of producing novel output which is valuable \cite{runco2012standard}, was proposed by Runco and Jaeger to facilitate the process of creativity evaluation with a workable definition. This definition captures two fundamentals of creativity, which can be treated as necessary - but not sufficient \cite{jordanous2016modelling}. Value is context dependent, as is novelty, and what is important for creativity in one type of creativity is not necessarily important for creativity in another area. As Plucker and Beghetto highlight, although creativity is essentially domain-general (fundamentally the same across all types of creativity), in practical terms we need to add domain-specific context to how creativity is assessed: 

\begin{quote}
`Creativity is the interplay between ability and process by which an individual or group produces an outcome or product that is both novel and useful as defined within some social context.'
\cite[p. 156]{plucker04}
\end{quote}

What social context should we take into account when evaluating creativity in music improvisation? Improvisation is the generation of music in real-time, and can be considered as the live demonstration of musical creativity: a creative process where the outputs are closely tied to the live process. In fact, improvisation and creativity are often conflated by authors rather than being distinct behaviours \cite{sawyer1999improvisation, thom2003interactive, johnson2002jazz, biasutti2009dimensions}.

While there have been several useful contributions on creativity in musical improvisation (e.g. \cite{johnson2002jazz, thom2003interactive, biasutti2009dimensions, jordanous12jims}), no overall consensus emerges for how that creativity is manifested. However one theme which emerges is the importance of emotional involvement and emotion-driven process \cite{biasutti2009dimensions, lewis2011, jordanous2016modelling}. Lewis coins improvisation as a process such that while improvising, ‘one hears something of oneself’ \cite{lewis2011}.

How can we model human emotions in a computational way, so computers can detect emotional content in musical content? While much work exists in music mood/emotion recognition \cite{han2022survey}, it can be argued that work in this area does not always align with psychologically-informed models of emotion. Two leading psychological models of emotion are Russell's circumplex model \cite{russell1980circumplex} and Ekman's model of basic emotions \cite{ekman1992argument}. Russell's two-dimensional model, treating valence (the positive/negative sentiment) and arousal (the amount of active energy perceived in a sound) as two axes by which all emotions can be plotted, has been shown useful for music emotion categorisation e.g. \cite {soleymani20131000, langroudi2018music}. However it does not capture the magnitude of emotions present, only the presence (or absence) of individual emotions detected in data. 

The Ekman model has been less explored in practical terms in music emotion research, probably due to the relative difficulty of detecting Ekman's basic emotions compared to the Russell model, which requires tests for two audio features (valence and arousal) that have been implemented computationally e.g. as evaluated in \cite{vidasvalidating}.  Perhaps somewhat overlooked, the work of Juslin and Laukka in identifiying and categorising acoustic cues for different types of emotion \cite{juslin03} collects together different audio-based cues for emotional content in audio (music performance and speech), through a review of 41 studies of musical performance and 104 studies of vocal expression. Juslin and Laukka's review identifies how acoustic cues from the reviewed studies fall into categories (anger, fear, happiness, sadness, tenderness) that align reasonably well with Ekman's model of basic emotions (anger, fear, happy, sad, disgust, and the later addition of surprise as a sixth basic emotion). The final category by Juslin and Laukka, tenderness, could be argued to be the antithesis of the disgust emotion named in Ekman's model. 

In contrast to the binary nature of emotion detection in Russell's model, the acoustic cues identified in Juslin and Laukka's review can be tested for in a way which allows us to measure the strength or magnitude of the presence of that acoustic cue, This allows us to go beyond categorisation of emotional content, to measure magnitudes at which different emotional indicators are present. It could also be argued that the Juslin and Laukka model is more grounded in psychological plausibility. Unlike many computational implementations of valence and arousal, which typically test for audio features that can be used as a proxy for the two features, the Juslin and Laukka model is the result of collating and combining multiple studies on acoustic cues for the presence of emotion in audio context. Given such psychological validation, this model seems useful for investigation in this work's goal: for detecting and comparing the presence and quantity of emotion-laden content. If the cues in Juslin and Laukka's model can be successfully implemented as tests which can be applied to audio at scale, then this gives us a psychologically validated set of metrics that can be applied to detect musical cues for emotion, at the scale needed for evaluation of current music generative AI tools. This provides the step forward for this current study: evaluating a contextually important factor of musical improvisation creativity using automated tests which can be applied at scale to large datasets.


\section{Methodology}

Audio data of various improvisations was collected for analysis. The goal was to have a dataset of more creative improvisations matched with a dataset of less creative improvisations. Ideally it would be helpful to have data on rankings of highly creative improvisations (or conversely, improvisations which are deemed to be of low creativity). To the best of my knowledge, such a dataset does not exist. For the purposes of this study, a working assumption is made that if we compare a set of notable improvisations by famous jazz musicians against a set of improvisations generated by an AI tool, or co-created by a human musician working with an AI tool, we can treat the notable improvisations as more creative than the co-created improvisations.\footnote{While this working assumption gives us an approach that allows us to proceed with data, it is perhaps somewhat unfair on the process involving the less notable musician and AI generator. It should be noted that the human musician involved, Bob Keller, was regarded as a competent musician.} The source of the co-created improvisations in this study was a set of tracks generated using the Impro-Visor music improvisation software \cite{keller2010}.

For a fair comparison between computer-generated and human-performed improvisations, it is important to detect the emotional content directly in the music data, as distinct from emotional content added during individual performance. Hence for this study, MIDI tracks generated by Impro-Visor were used for all audio data, being converted to a signal-based audio representation for more straightforward analysis. This removed any additional emotional indicators added by a human performer, allowing analysis to focus purely on the musical data. Data were collected from \url{https://www.cs.hmc.edu/~keller/jazz/improvisor/Solos/}, comparing transcriptions of improvisations by Charlie Parker, Clifford Brown and Dave Liebman with a set of improvisations generated either by Impro-Visor or by a human musician (Bob Keller) working with the Impro-Visor software. 

Juslin \& Laukka features were implemented to analyse audio data, as reported in the next section. As part of this implementation, a benchmark identification phase was necessary; each of the acoustic cues used by Juslin and Laukka rely on descriptive terms to distinguish between how musical features are used in each emotional category. For example, the anger, fear and happiness emotions are categorised in \cite{juslin03} as typically associated with `fast' tempos, with the sadness and tenderness emotions typically associated with `slow' tempos. Hence a set of emovectors was calculated for a benchmark dataset (the jazz subset of the GTZAN dataset \cite{tzanetakis2002musical}), to provide percentiles as thresholds for each acoustic cue as per Table \ref{tab:jlthresholds}.  

With the calculated thresholds, the collected improvisations data were analysed to judge differences in the acoustic cues being measured. The output of the analysis process are \emph{emovectors}: 5-dimensional embeddings or vectors, one for each song in analysis data. The five dimensions represent the five emotions categorised by Juslin and Laukka: anger, fear, happiness, sadness and tenderness. 

Emovector results across sets of improvisations were compared across the two sets of improvisations collated for comparison, to investigate the hypothesis that more creative improvisations contain more emotion-laden content.

\section{Operationalising the acoustic cues}


The acoustic cues identified by Juslin and Laukka were compiled into Table \ref{tab:jlcues}, representing how each acoustic cue differed per emotion.  
\begin{table*}
    \centering
    \begin{tabular}{c | c c c c c c c c c }
    & Tempo&	sound &	sound level &high-frequency &pitch 	&pitch &pitch &tone &microstructural \\
    & &	 level& variability	& energy	& level	& variability	& contour	& attacks	& irregularity \\
    \hline
Anger	&Fast &	high 	&much &	much &	high 	&much 	&rising 	&fast 	&Irregularity \\
Fear	&Fast &	low 	&much &	little 	&high &	little 	&rising 	&n/a	&a lot of irregularity\\
Happiness	&Fast 	&medium–high &	n/a	&medium &	high &	much &	rising &	fast &	very little regularity\\
Sadness 	&Slow 	&low &	little &	little 	&low 	&little 	&falling &	slow &	irregularity\\
Tenderness	&Slow 	&low 	&little 	&little 	&low &	little 	&falling &	slow 	&regularity\\
\end{tabular}
    \caption{Mapping of acoustic cues to emotion.}
    \label{tab:jlcues}
\end{table*}
The Table \ref{tab:jlcues} were then converted into a set of indicative measurements per acoustic cue for each emotion, representing concepts such as `high', `little' etc in a numeric form, as represented in Table \ref{tab:jlthresholds}.

\begin{table*}
    \centering
    \begin{tabular}{c|ccccccccc }
    & Tempo&	sound &	sound level &high-frequency &pitch 	&pitch &pitch &tone &microstructural \\
    & &	 level& variability	& energy	& level	& variability	& contour	& attacks	& irregularity \\
    
    \hline
Anger	&1	&1	&1&	1	&1&	1	&1&	1	&0.5 \\
Fear	&1	&0&	1	&0	&1	&0	&1	& n/a &	1\\
Happiness&	1	&0.75	& n/a	&0.5	&1	&1	&1	&1	&0.25\\
Sadness 	&0&	0	&0	&0&	0&	0	&0	&0&	0.5\\
Tenderness&	0&	0&	0&	0	&0	&0	&0	&0&	0\\
\end{tabular}
    \caption{Representation of acoustic cue as a set of indicative measurements per acoustic cue for each emotion, in the interval [0, 1]. }
    \label{tab:jlthresholds}
\end{table*}
	
Evaluation of each of the acoustic cues was operationalised as follows, using the Librosa \cite{mcfee2015librosa} Python library (see \url{https://github.com/annajordanous/emovectors} for code and data).

 \begin{itemize}
 \item tempo: calculated using Librosa's beat\_track() function.
 \item microstructural irregularity: treated as beat irregularity, using Librosa's frames\_to\_time() function to convert beats to a time-based representation, then calculating the extent to which each beat deviates from expected tempo.
 \item sound level: calculating the root-mean-square (RMS) value for each frame (sample segment) of the audio using Librosa's rms() function, to represent the amount of energy present in the audio signal at each frame.
 \item sound level variability: using the RMS calculations for sound level per frame and calculating the extent to which frames differed from each other over the entire audio.
 \item high-frequency energy: calculating the mean spectral bandwidth for each audio; the wider (larger) the bandwidth on average, the more high-frequency energy is present, as the bandwidth represents the span from highest to lowest frequencies present in a sound signal.
 \item pitch level: calculated using Librosa's yin() function, which calculates the fundamental frequency of the sound signal (the lowest frequency present) for each phrase.
 \item pitch variability: calculating the extent to which the pitch level differs across all frames of the audio.
 \item tone attacks: calculating attack times (the initial part of a sound `envelope', which represents how sound changes over time for individual sounds such as a note being played) by using the Librosa rms() function to identify frames with levels of energy indicating an attack (implemented as above a threshold of 10\% of the maximum energy detected in the audio) and calculating the number of frames for which an attack lasts.
 \end{itemize}

One acoustic cue has not yet been implemented, which is the pitch contour cue. This is complex to calculate for polyphonic audio which contains several instruments. In future work, this could be implemented if working with monophonic data such as a solo played by an instrument such as a saxophone or trumpet, which typically can only sound one note at a time.



\section{Results and Discussion}

\begin{table}[]
    \centering
    \begin{tabular}{c|c c}
         & Transcribed famous 
&  Impro-visor/Keller 
\\
                 & solos (n=6)
&   solos (n=11)
\\ 
\hline
Anger &
mean = 2.000
s.d = 0.894 &
mean = 1.545 
s.d = 1.128 \\
Fear&
mean = 2.000
s.d = 1.095 &
mean = 1.364
s.d = 0.674 \\
Happiness&
mean = 1.333
s.d = 1.506&
mean = 1.091
s.d = 0.831\\
Sadness&
mean = 2.000
s.d = 1.897&
mean = 1.727
s.d = 1.104\\
Tenderness&
mean = 2.000
s.d = 1.549&
mean = 1.546
s.d = 1.036\\
    \end{tabular}    
    \caption{Comparative results between each dataset}
    \label{tab:comparison}
\end{table}

Results, as presented in Table \ref{tab:comparison}, showed that for each of the five emotions, on average (mean), each emotion was present to a greater extent than in the Impro-visor/Keller solos. While the datasets are too small for analysis of significance of results, this represents promising initial results for our hypothesis that if an improvisation contains more evidence of emotion-laden content, it is more likely to be recognised as creative. 

Operationalising these tests give us an approach that can be run at scale, with no human interaction required. This is particularly relevant when wishing to analyse the (often vast) output of an AI generative music tool. 

\section{Future work}

As mentioned above, one acoustic cue will be implemented in future work (pitch contour) with different data. Once larger datasets can be sourced for analysis, the evaluations can be rerun at larger scale, and this will allow for statistical analysis across songs matched to ranking to verify hypothesis. What we lack is any dataset where improvisations are somehow ranked for creativity; this represents a significant area for future data collection activity. Such a dataset would be a valuable resource for future work on music improvisation creativity. 

Once tests can be implemented at scale and validated against other rankings of creativity, the hypothesis of this work can be fully evaluated. If successful, this method enables investigations such as analysis of how chord sequences influence creativity potential for improvisation, or analysis of creativity of an individual artist across different recordings, time periods or musical groups they play in.

\section{Conclusions}

Demonstration of emotional involvement is closely linked with creativity in music improvisation. This study hypothesises that if an improvisation contains more evidence of emotion-laden content, it is more likely to be recognised as creative. This is difficult to test for in current LLM-based generative systems or for large music data sets, as automated metrics for emotion detection typically either require human interaction, which does not scale up for large datasets, or may be somewhat distinct from psychologically validated models of emotion. 

The results of this study thus far are based on small datasets and with a number of limitations based on the lack of validated data ranking creativity of improvisations. Initial results do show promising results for the theory that higher emotional content in music can be used to contribute towards new metrics for creativity evaluation that can be applied at scale.

\section{Ethics statement}
No experiments were conducted with human participants for this study.  Training data were from licensed open data sources.

\section{Acknowledgements}
For the purpose of open access, the author has applied a Creative Commons Attribution (CC BY) licence to any Author Accepted Manuscript version arising.

\bibliographystyle{ieeetr}
\bibliography{references}


\end{document}